\documentclass[aps,prl,preprint,   superscriptaddress,longbibliography]{revtex4-2}
\usepackage{graphicx}
\usepackage{braket}
\usepackage{amsmath}
\usepackage{amssymb}
\usepackage{mathrsfs}
\usepackage{amsmath}
\usepackage{color}
\usepackage{epsfig}

\usepackage[version=3]{mhchem} % Formula subscripts using \ce{}
\usepackage[T1]{fontenc}       % Use modern font encodings
\usepackage{amsmath}
\usepackage{amssymb}
\usepackage[export]{adjustbox}[2011/08/13]
\usepackage{bm}
\usepackage{ulem}

\usepackage{xr}
\makeatletter
\newcommand*{\addFileDependency}[1]{
  \typeout{(#1)}
  \@addtofilelist{#1}
  \IfFileExists{#1}{}{\typeout{No file #1.}}
}
\makeatother

\newcommand*{\myexternaldocument}[1]{
    \externaldocument{#1}
    \addFileDependency{#1.tex}
    \addFileDependency{#1.aux}
}
\myexternaldocument{HHGBi2Se3_SM}

\usepackage{hyperref}
\hypersetup{
     colorlinks   = true,
}
\begin{document}

\title{All-optical probe of three-dimensional topological insulators based on high-harmonic generation by circularly-polarized laser fields} 
%% Each author should be given as a separate \author command.
%%
%% Corresponding authors should have an e-mail given after the author
%% name as an \email command. Phone and fax numbers can be given
%% using \phone and \fax, respectively; this information is optional.
%%
%% The affiliation of authors is given after the authors; each
%% \affiliation command applies to all preceding authors not already
%% assigned an affiliation.
%%
%% The affiliation takes an option argument for the short name.  This
%% will typically be something like "University of Somewhere".
%%
%% The \altaffiliation macro should be used for new address, etc.
%% On the other hand, \alsoaffiliation is used on a per author basis
%% when authors are associated with multiple institutions.
%%%%%%%%%%%%%%%%%%%%%%%%%%%%%%%%%%%%%%%%%%%%%%%%%%%%%%%%%%%%%%%%%%%%%
\author{Denitsa Baykusheva}
\email{denitsab@stanford.edu}
\affiliation{Stanford PULSE Institute, SLAC National Accelerator Laboratory, Menlo Park, California 94025, USA}

\author{Alexis Chac\'{o}n}
\affiliation{Center for Nonlinear Studies and Theoretical Division, Los Alamos National Laboratory, Los Alamos, New Mexico 87545, USA}
\affiliation{Department of Physics and Center for Attosecond Science and Technology, POSTECH, 7 Pohang 37673, South Korea}
\affiliation{Max Planck POSTECH/KOREA Research Initiative, Pohang 37673, South Korea}

\author{Jian Lu}
\altaffiliation{Present address: Department of Physics and Astronomy, University of Pennsylvania, 209 South 33rd Street, Philadelphia, PA 19104-6396, USA}
\affiliation{Stanford PULSE Institute, SLAC National Accelerator Laboratory, Menlo Park, California 94025, USA}

\author{Trevor P. Bailey}
\affiliation{Department of Physics, University of Michigan, Ann Arbor, Michigan 48109, USA}

\author{Jonathan A. Sobota}
\affiliation{Stanford Institute for Materials and Energy Sciences, SLAC National Accelerator Laboratory, Menlo Park, California 94025, USA}

\author{Hadas Soifer}
\altaffiliation[Present address: ]{Raymond and Beverly Sackler School of Physics and Astronomy, Tel Aviv University, Tel Aviv 69978, Israel}
\affiliation{Stanford Institute for Materials and Energy Sciences, SLAC National Accelerator Laboratory, Menlo Park, California 94025, USA}

\author{Patrick S. Kirchmann}
\affiliation{Stanford Institute for Materials and Energy Sciences, SLAC National Accelerator Laboratory, Menlo Park, California 94025, USA}

\author{Costel R. Rotundu}
\affiliation{Stanford Institute for Materials and Energy Sciences, SLAC National Accelerator Laboratory, Menlo Park, California 94025, USA}

\author{Ctirad Uher}
\affiliation{Department of Physics, University of Michigan, Ann Arbor, Michigan 48109, USA}

\author{Tony F. Heinz}
\affiliation{Stanford PULSE Institute, SLAC National Accelerator Laboratory, Menlo Park, California 94025, USA}

\author{David A. Reis}
\affiliation{Stanford PULSE Institute, SLAC National Accelerator Laboratory, Menlo Park, California 94025, USA}

\author{Shambhu Ghimire}
\email{shambhu@slac.stanford.edu}
\affiliation{Stanford PULSE Institute, SLAC National Accelerator Laboratory, Menlo Park, California 94025, USA}

\date{\today}

\begin{abstract}

We report the observation of a novel nonlinear optical response from the prototypical three-dimensional topological insulator Bi$_2$Se$_3$ through the process of high-order harmonic generation. We find that the generation efficiency increases as the laser polarization is changed from linear to elliptical, and it becomes  maximum for circular polarization. With the aid of a microscopic theory and a detailed analysis of the measured spectra, we reveal that such anomalous enhancement encodes the characteristic topology of the band structure that originates from the interplay of strong spin-orbit coupling and time-reversal symmetry protection. Our study reveals a new platform for chiral strong-field physics and presents a novel, contact-free, all-optical approach for the spectroscopy of topological insulators. The implications are in ultrafast probing of topological phase transitions, light-field driven dissipationless electronics, and quantum computation.

%we find that the no Topological insulators exhibit a unique combination of conducting surface states protected by time-reversal symmetry and insulating bulk bands, thus endowing them with novel opportunities for both fundamental science and technological advances. The ultrafast optical response of these materials to intense light fields could reveal their full potential, including that present in transient non-equilibrium states. Here, we report the observation of non-perturbative high-order harmonic generation from a topological insulator, Bi$_2$Se$_3$, subjected to strong mid-infrared laser fields. 

\end{abstract}

%\keywords{}

\maketitle

\section*{Introduction}

High-harmonic spectroscopy is an all-optical method to probe the structure and ultrafast dynamics of atoms and molecules subjected to strong laser fields~\cite{Itatani2004,Peng2019}. The basic idea is based on the underlying microscopic mechanism of the high harmonic generation (HHG) process, which can be largely understood with the aid of a three-step recollision model~\cite{Kulander1993,Corkum1993} comprising sub-cycle tunneling of electrons, their  laser-field acceleration in the vacuum, and their subsequent returns with possible recombination to parent ions. The recombination is accompanied by the emission of high-order harmonics whose amplitude and phase depends on the recollision angle and time, providing HHG with an atomic-scale spatial and temporal resolution ~\cite{nirit2020}. Analogous processes are also being investigated in solid-state HHG ~\cite{Ghimire2011,Vampa2015a,Luu2015,Rubio2018,Ghimire2018}, leading to the possibility of compact attosecond light sources ~\cite{Ndabashimiye2016,Garg2018,Li2020,hamed2021} and to extend these all-optical spectroscopic abilities to solid materials such as semiconductors ~\cite{Vampa2015}, dielectrics ~\cite{Luu2015} and to a range of quantum materials (QMs). Recent experimental efforts on high-harmonic spectroscopy of QMs include the observation of non-perturbative high-harmonics from graphene \cite{Yoshikawa2017}, monolayer transition metal dicacogenides (TMDC) ~\cite{Liu2017a}, and topological insulators (TIs) \cite{Baykusheva2019,Bai2020,Schmid2021} subjected to intense laser fields without damage.

Three-dimensional TIs (3D TIs) are characterized by the presence of insulating bulk bands and conducting surface bands forming an odd number of Dirac cones within the bulk band gap as a result of band inversion owing to the strong spin-orbit coupling~\cite{Hasan2010,Qi2011}. Due to a multitude of emergent phenomena, ranging from the unique spin texture~\cite{Hsieh2009} supporting helical currents~\cite{McIver2012} to dissipationless transport~\cite{Roushan2009}, TIs are also anticipated to serve as building blocks for future technology. In particular TIs may host the fastest optical transients \cite{OliaeiMotlagh2018}, and have applications in spintronics, in quantum sensing and metrology, as well as in quantum computation \cite{Moore2010}. Therefore, it is desirable to develop advanced methods that can probe TIs including their non-equilibrium states. Currently, scanning tunnelling microsocpy (STM) and angle-resolved photoemission spectroscopy (ARPES) \cite{Xia2009} are the most common methods to probe TIs, although they require stringent sample environments and are not compatible with the use of strong light-fields. Therefore, the development of purely optical methods that are endowed with the required temporal resolution and are compatible with a variety of sample environments including external magnetic fields~\cite{Wu2016}  and ambient conditions is highly desired. 

To this aim, in addition to the demonstration of symmetry breaking at the surface through the measurement of even-order harmonics \cite{Baykusheva2019,Bai2020,Schmid2021}, it is crucial to elucidate the imprint of characteristic features of 2D TIs on the high harmonic spectra, particularly the strong spin-orbit coupling (SOC). The SOC gives rise to the band inversion and  the crossing of the surface valence and conduction bands at or near the Fermi level. In a recent theoretical work, we predicted a unique enhancement of high harmonics for a circularly-polarized driving laser fields from the surface state of a prototypical topological insulator system Bi$_2$Se$_3$ \cite{Baykusheva2021}. Here, we report experimental results. We use sub-band-gap, ultrashort, intense mid-infrared (MIR) laser pulses to  induce non-perturbative high-harmonic generation in ultra-thin single-crystals using transmission geometry, as well as in freshly-cleaved bulk crystals in reflection mode by avoiding material's self absorption of high-energy photons. First, we discuss results from linearly polarized laser fields, and analyse the intensity and polarization of high harmonics with respect to the crystal orientation. This allows us to identify spectral features that we attribute to the contribution from the surface. Then, we present main results, which include the anomalous high-harmonic response of the topological material to circularly polarized laser fields.

\section*{Results and Discussion}

\subsection*{Non-perturbative high-harmonic generation driven by ultrashort MIR pulses}

\begin{figure}[htb!]
\begin{center}
\includegraphics[scale=0.4]{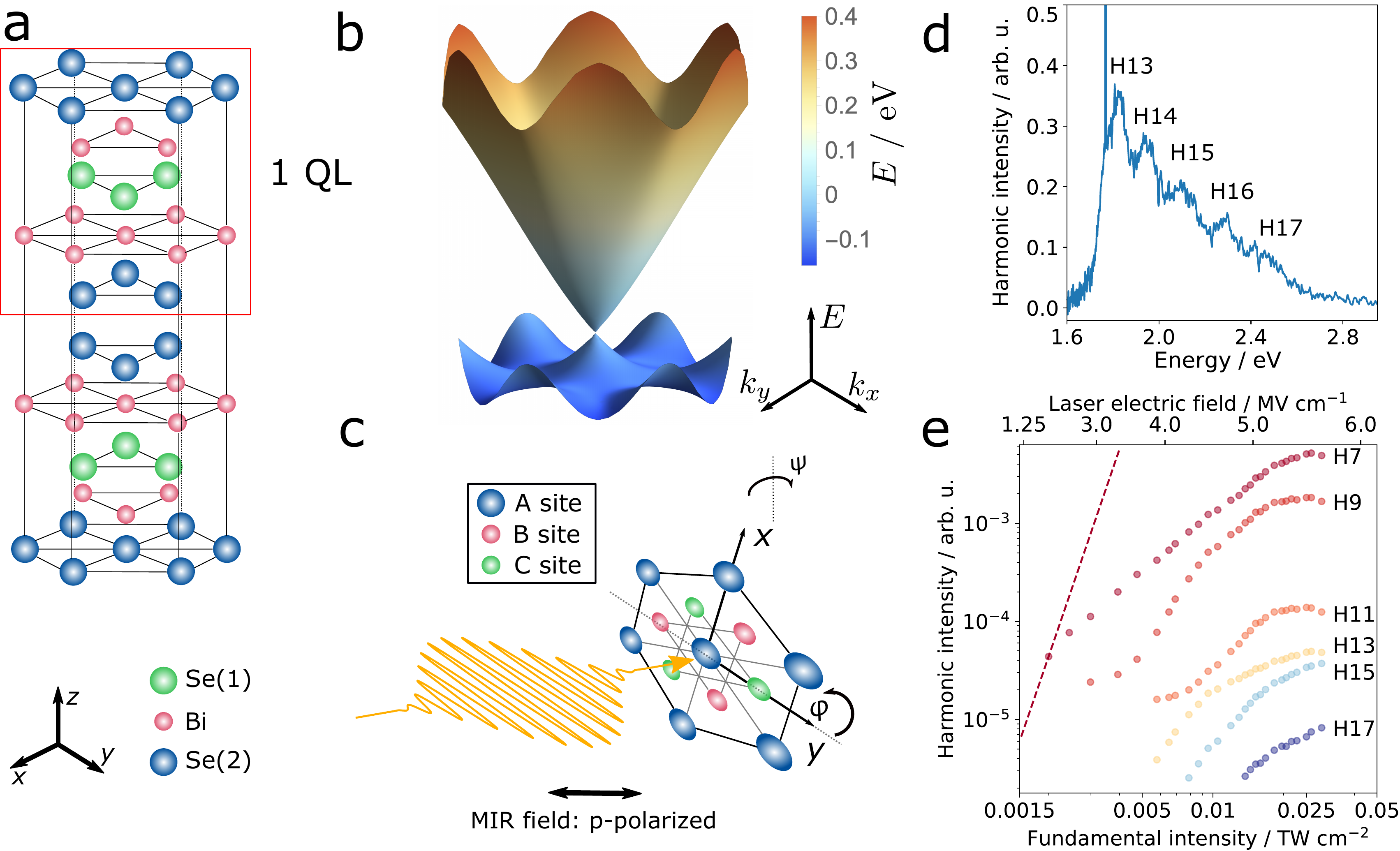}
\end{center}
\caption{\textbf{High harmonics from Bi$_2$Se$_3$}: (a): Real space crystal structure Bi$_2$Se$_3$. The crystal is composed of alternating plane layers of Bi and Se atoms, aligned along the $z$-direction. The building unit of the lattice is formed by quintuple layers (QL). Each QL features five atoms: 2 symmetry-equivalent Bi sites, 2 equivalent Se sites (Se(1)), and a third Se atom (Se(2), which coincides with the inversion center). (b): 3D-representation of the energy dispersion ($k_\parallel\le0.4$~\AA$^{-1}$) of the surface states. (c): Schematics of a crystal-orientation dependence study, where HHG yield is recorded as a function of the azimuthal angle $\varphi$, defined as the offset between the mirror plane of the crystal (dashed line) and the plane of incidence. (d): Representative harmonic spectrum from a bulk sample, recorded in reflection geometry ($\lambda_\mathrm{MIR}=8.7$~$\mathrm{\mu m}$). The peak marked with an asterisk is an artefact caused by stray light. Due to the limitations in our detection system, lower-order harmonics are not accessible. (e): Intensity dependence of the HHG yield from  a  thin film sample in transmission geometry ($\lambda_\mathrm{MIR}=5.2$~$\mathrm{\mu m}$). The yellow dashed line indicates the power dependence of HO~7 predicted by the perturbative scaling law. Crystal structure schemes are adopted from ~\cite{Baykusheva2021}.}
\label{fig1}
\end{figure}

Bi$_2$Se$_3$ (space group $R\overline{3}m$, $D_{3d}^5$, $\sharp$166)~\cite{Zhang2009} has a layered rhombohedral structure composed of covalently bonded quintuple layers (QLs) stacked along the trigonal axis and stabilized by weak van der Waals interactions (Fig.~\ref{fig1}~a). The bulk has a moderately large band gap ($\sim0.3$~eV), which makes it suitable for non-resonant, below-band-gap excitation in the MIR wavelength range. Our experimental setup, outlined in detail in the Materials and Methods section below as well as in Sec.~\ref{sec:setup} in the SM, allows for the generation of highly tunable MIR laser pulses spanning the range between 5 and 10~$\mathrm{\mu m}$  ({0.12} -- 0.25~eV). The MIR laser is focused  to a  spot size of $\sim300\times300$~$\mu$m$^2$, yielding vacuum peak intensities of up to $I_0\sim 0.03$~\mbox{TW/cm${}^2$} (field strength in the sample $\sim0.03$~\mbox{V/\AA}), see  supplementary material (SM) for details. 

The band structure of the surface states (Fig.~\ref{fig1}~b) features a single Dirac point characterized by nearly linear dispersion. In our experiments, we generate high harmonics  in transmission, at normal incidence, from thin films of sub--100-nm thickness deposited on a (111)-oriented, optically isotropic BaF$_2$-substrates. 
To confirm that the emission is from the pristine sample, we also perform similar measurements in freshly cleaved (in air) bulk Bi$_2$Se$_3$, but using specular-reflection-geometry to avoid absorption losses (see Materials and Methods section). A typical MIR-driven HHG spectrum comprising both even and odd harmonic orders 
is shown in panel d of Fig. \ref{fig1}, displaying distinct harmonic peaks spanning the range from above-band-gap harmonic order (HO) 13 ($\sim 1.85$~eV) to 17 ($\sim 2.41$~eV). The experimentally determined dependence of the harmonic yield  on the driving field intensity  presented in Fig.~\ref{fig1}~e clearly demonstrates the non-perturbative response. All harmonic orders exhibit a  behavior markedly different from the predictions of the perturbative response ($I_\mathrm{HHG}\propto I_0^h$, whereby $h$ is the harmonic order).

\subsection*{Polarization-resolved analysis of the high-harmonic response}

Bulk Bi$_2$Se$_3$ is centrosymmetric, and exhibits a three-fold rotational symmetry ($C_3$) along the $z$-axis. However, at the surface ($xy$-plane in Fig.~\ref{fig1}~c, $P3m1$ wallpaper group ($C_{3v}$)), inversion symmetry is necessarily broken and there is a mirror axis along the $y$-direction. This distinction allows us to attribute the observed even-harmonic generation (cp. Fig.~\ref{fig1}~d as well as Fig.~\ref{fig:orient_integ} in the SM) to the participation of surface electronic states in the HHG process. In what follows, we  concentrate our analysis on the three representative harmonic orders:  HO~7, 8 and 9 of $7.5$~$\mathrm{\mu m}$ (see SM, Fig.~\ref{fig:orient_integ} for spectra).

\begin{figure}[htb!]
\begin{center}
\includegraphics[scale=0.5]{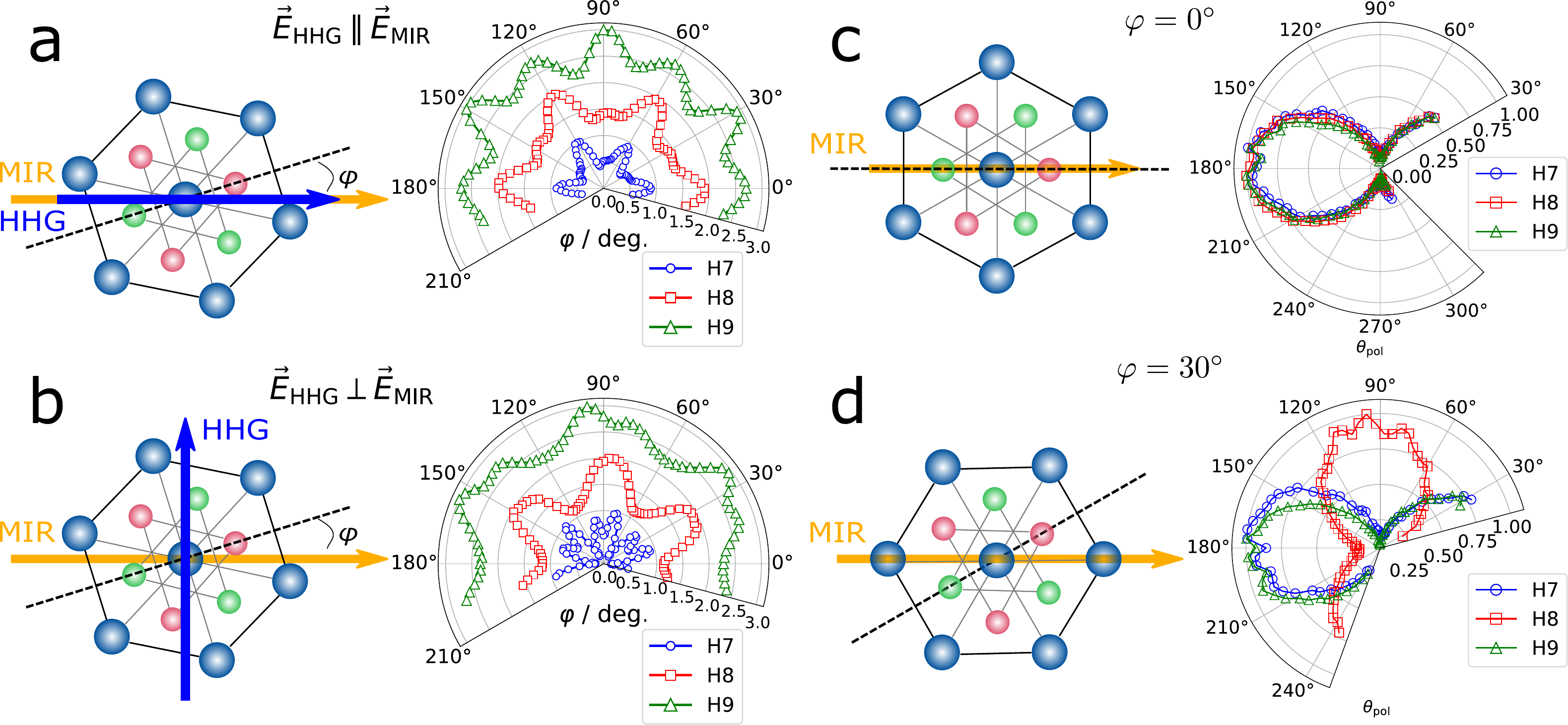}
\end{center}
\caption{\textbf{HHG orientation and polarization analysis}: Measured crystal-orientation dependent HHG yields for HOs 7, 8, and 9 for fixed polarization anlges, $\theta_\mathrm{pol}$, along (a) and perpendicular (b) to the polarization direction of MIR laser field.  The relative orientation between the MIR laser polarization (horizontal) and the crystal's mirror axis (dashed line) is quantified by $\varphi$, as shown schematically on the left side in each panel. The curves for various harmonics are separated by an offset of 1 for clarity. Polarization measurement of HOs 7, 8, and 9 for MIR laser field along (c) and perpendicular (d) to the crystal's mirror axis. The non-vanishing signal for HO~8 at $\theta_\mathrm{pol}=0^\circ$ in panel~d is due to spectral contamination from the strong  HO~7 and HO~9.}\label{fig2}
\end{figure}

Panels a and b of Figure~\ref{fig2} show the crystal-orientation-dependent HHG yield ($I_\mathrm{HHG}$), disentangled in components parallel ($\vec{E}_\mathrm{HHG}\parallel\vec{E}_\mathrm{MIR}$) and perpendicular ($\vec{E}_\mathrm{HHG}\perp\vec{E}_\mathrm{MIR}$) with respect to the driving laser field ($\vec{E}_\mathrm{MIR}$). The three-fold crystal symmetry is reflected in the six-fold periodicity of the HHG signal as a function of the azimuthal angle $\varphi$ defined in Fig.~\ref{fig1}~c. The parallel component (panel~a) has local maxima along ($\varphi=0^\circ,  60^\circ \ldots$) and at $30^\circ$  $(\varphi=30^\circ, 90^\circ\ldots)$ with respect to the mirror axis. We note that in the reduced Brilloiuin zone (BZ) of the surface, these two directions correspond to the  high-symmetry directions $\overline{\Gamma M}$ and $\overline{\Gamma K}$, respectively (s. Fig.~\ref{fig:BZ}~a in the SM). 
In contrast, the perpendicular ($\vec{E}_\mathrm{HHG}\perp\vec{E}_\mathrm{MIR}$) components exhibit minima along the mirror axes. \\

Panels c and d of Fig.~\ref{fig2} display the polarization ellipses of HOs 7 to 9 when the mirror axis is along ($\varphi=0^\circ$) and perpendicular ($\varphi=30^\circ$) to the laser field, respectively. It is seen that in both cases, the polarization of the odd orders follows the driving field ($\vec{E}_\mathrm{HHG}^\mathrm{odd}\parallel\vec{E}_\mathrm{MIR}$), whereas that of the even order (HO~8) depends on the orientation: it is parallel to the laser field vector  when the latter is fixed at $\varphi=0^\circ$, and perpendicular to  it for the case $\varphi=30^\circ$. These results can be rationalized with the aid of symmetry arguments as introduced in Ref.~\cite{Neufeld2019} and elaborated for the case of Bi$_2$Se$_3$ in Ref.~\cite{Baykusheva2021}.  We note that perpendicularly-polarized even orders have been reported in other non-centrosymmetric materials such as $\alpha$-quartz~\cite{Luu2018} and polytype $\varepsilon$-GaSe~\cite{Langer2017} but the special aspect  regarding Bi$_2$Se$_3$ is that the inversion symmetry is broken only at the surface ~\cite{Hsieh2011b, Bai2020}.

\begin{figure}[htb!]
\begin{center}
\includegraphics[scale=0.5]{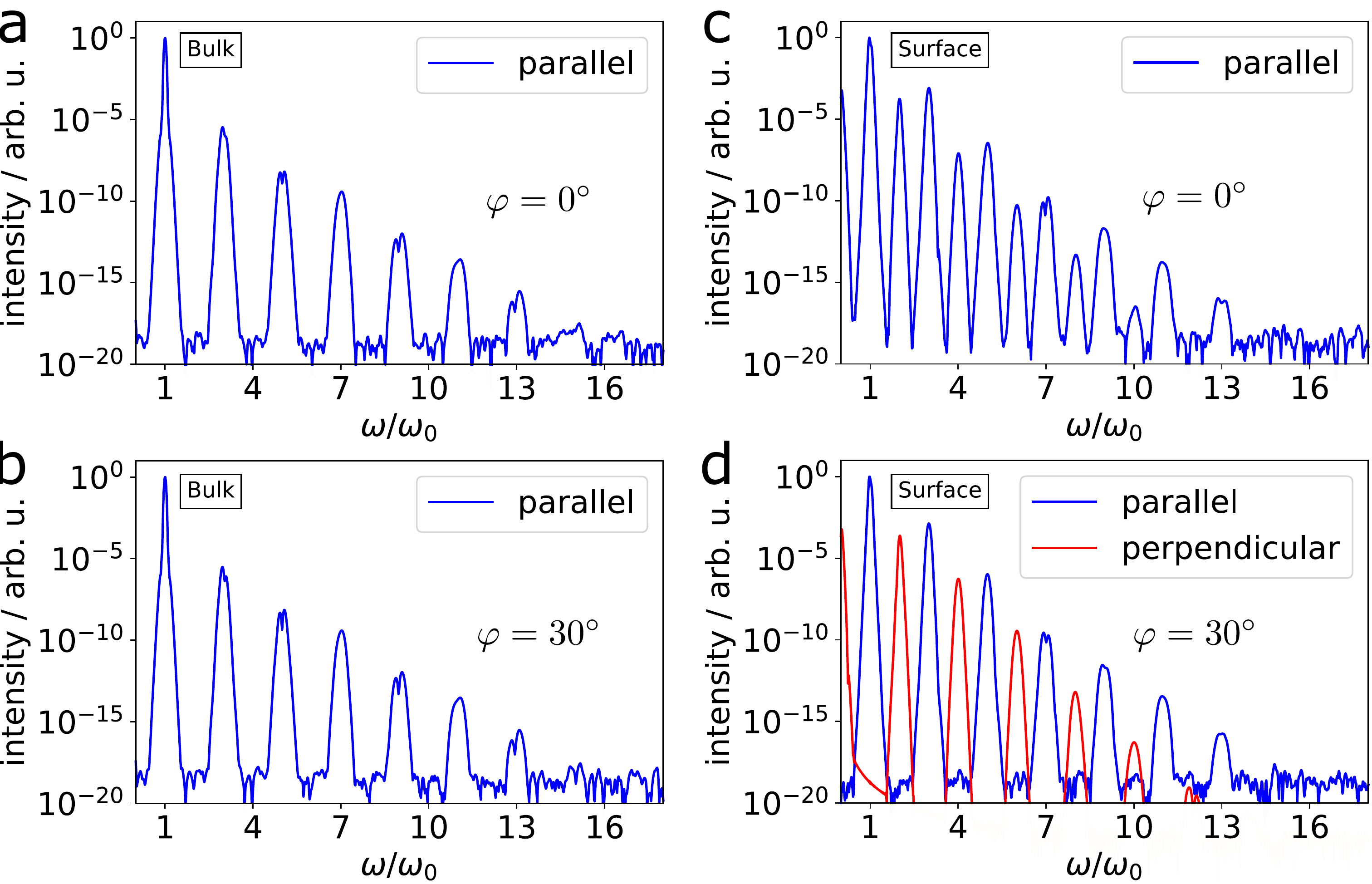}
\end{center}
\caption{\textbf{Calculated HHG response}: (a), (b): Calculated HHG spectrum for the bulk system subject to a linearly polarized MIR pulse with a center wavelength of $\lambda_\mathrm{MIR}=7.5$~$\mu$m and comprising 12 optical cycles  with an intensity of $I_0=0.0025$~TW/cm$^2$. The polarization vector is parallel to the mirror plane ($\varphi=0^\circ$) in panel (a) and orthogonal to it in panel (b) ($\varphi=30^\circ$). Panels (c) and (d) show an analogous calculation for the surface state, whereby the perpendicular component is shown in red.} \label{fig3}
\end{figure}

For comparison, we show calculated HHG spectra for a linearly polarized laser field oriented along ($\varphi=0^\circ$) and perpendicular ($\varphi=30^\circ$) to the mirror plane, separately for the bulk and the surface states in panels a-d of Fig.~\ref{fig3}. These calculations follow the standard semiconductor Bloch equations (SBE) approach ~\cite{Kira2011} and include geometrical effects, such as (non-Abelian) Berry curvature and Berry connections that are particularly important in the presence of gapless Dirac bands. The band structure was obtained using $4\times4$-tight-binding-model (TBM) Hamiltonian formulated in Ref.~\cite{Mao2011} and the topological surface states (TSSs) were deduced by applying open boundary conditions~\cite{Shan2010, Liu2010}. The detail calculation results and theoretical methodology can be found in Ref.~\cite{Baykusheva2021}. It can be seen that bulk states produce odd-order harmonics only, while surface state produce both odd and even-order harmonics. 
 
In particular, when the laser field is oriented along the $\varphi=30^\circ$ direction (orthogonal to the mirror plane), the orthogonally polarized even-order harmonics become significantly more intense. The odd orders in the direction orthogonal to the driving field vanish for both configurations. These results capture the experimental observations discussed above on a qualitative level. The rest of the manuscript  is devoted to the novel ellipticity-dependent response of the material and its connection to the specific features in the band structure. For this aim, we study how the harmonic intensity varies with the polarization state of the driving MIR laser field as the latter is changed from linear to elliptical and fully circular.

\subsection*{Non-trivial ellipticity response}

\begin{figure}[htb!]
\begin{center}
\includegraphics[width=\textwidth]{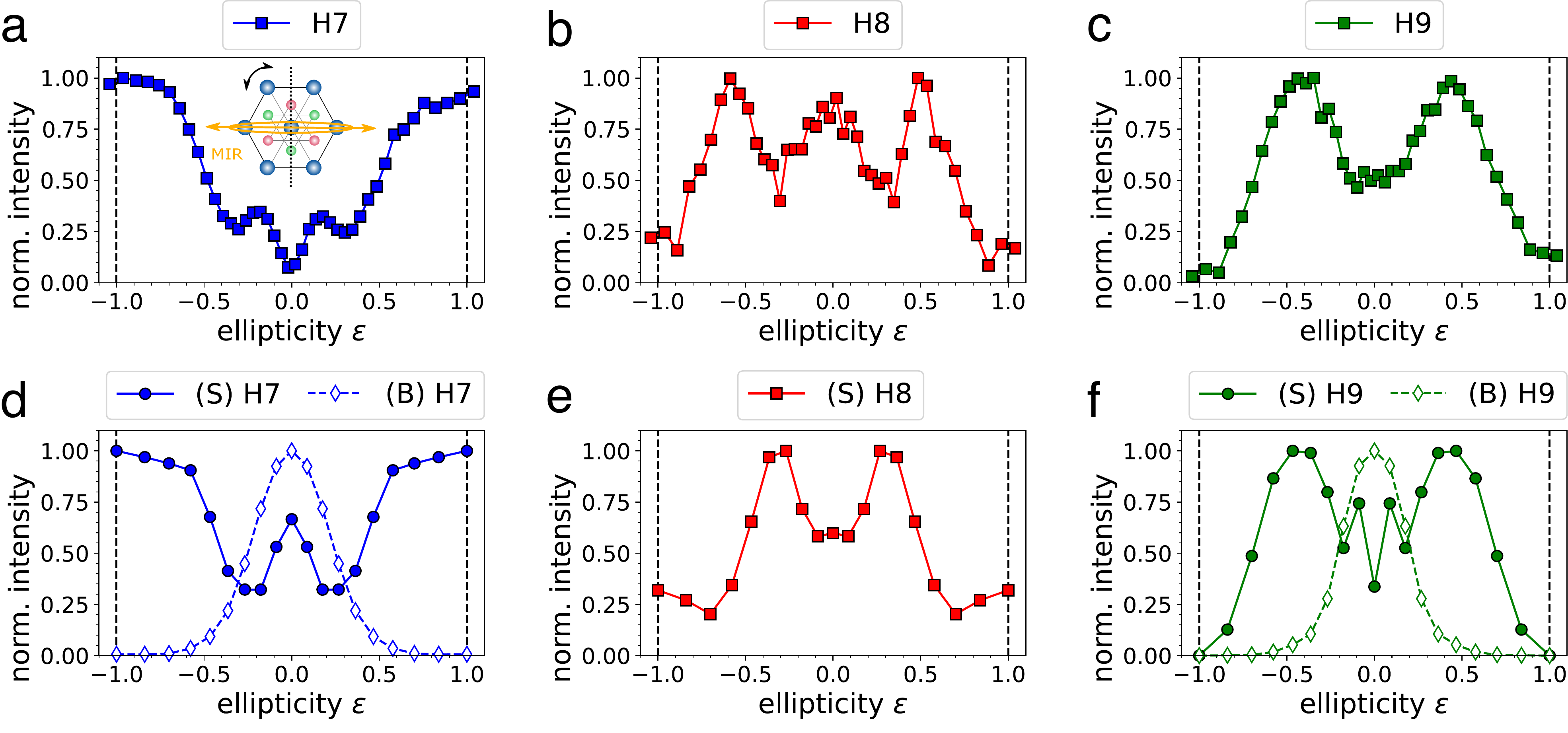}
\end{center}
\caption{\textbf{HHG ellipticity dependence}: 
(a), (b), (c): Experimental results; ellipticity dependent response of HHG measured at normal incidence using an ultrathin  sample.  The inset shows the experimental geometry, where the major axis of the MIR  ellipse (indicated by the orange arrow) is kept  orthogonal to the mirror axis (dashed black line) of the crystal throughout the measurement. (d), (e), (f): Calculated ellipticity response (for a 12-cycle Gaussian pulse, $I_0=0.0075$~TW/cm$^2$ at $\epsilon=0$). The contributions of the bulk (``B'', dashed lines, empty symbols) and the the surface (``S'', solid lines, filled symbols) are shown separately. Crystal symmetry prohibits HO 9 for circular polarization. }\label{fig4}
\end{figure}

We study the response to the ellipticity of the laser field  by varying the polarization state of the laser from right-circularly polarized (RCP, $\epsilon=+1$) through linear (LIN, $\epsilon=0$) to left-circularly polarized (LCP, $\epsilon=-1$) in small increments. We keep the major polarization axis fixed along $\varphi=30^\circ$ throughout the scan. Figure~\ref{fig4} shows representative results, and we refer to the supplementary information section (Sec.~\ref{sec:ell}) for additional data sets. Evidently, the intensity of HO~7 increases as the ellipticity increases from $\epsilon=0$ to $\epsilon=\pm1$, in stark contrast to the response of isotropic gas media or other solid-state systems~\cite{Ghimire2011, Ndabashimiye2016, Liu2017a}, where the harmonic signal drops at high ellipticities. Additionally, HO~7 (cp. Fig.~\ref{fig:ell_supp} of the SM) has the same handedness as the fundamental circularly polarized laser field, and so does HO~13, which is in accordance with the selection rules for HHG in circularly polarized fields (cp. Ref.~\cite{Neufeld2019}). Both HO~7 and HO~13 show strong enhancement, approaching an order of magnitude ($I_\mathrm{HHG}^\mathrm{(CPL)}:I_\mathrm{HHG}^\mathrm{(LIN)} \approx 6$ for HO~7 and $I_\mathrm{HHG}^\mathrm{(CPL)}:I_\mathrm{HHG}^\mathrm{(LIN)} \approx 10$ for HO~13), compared to the linear polarization for the same peak intensity.

\begin{figure}[htb!]
\begin{center}
\includegraphics[scale=0.365]{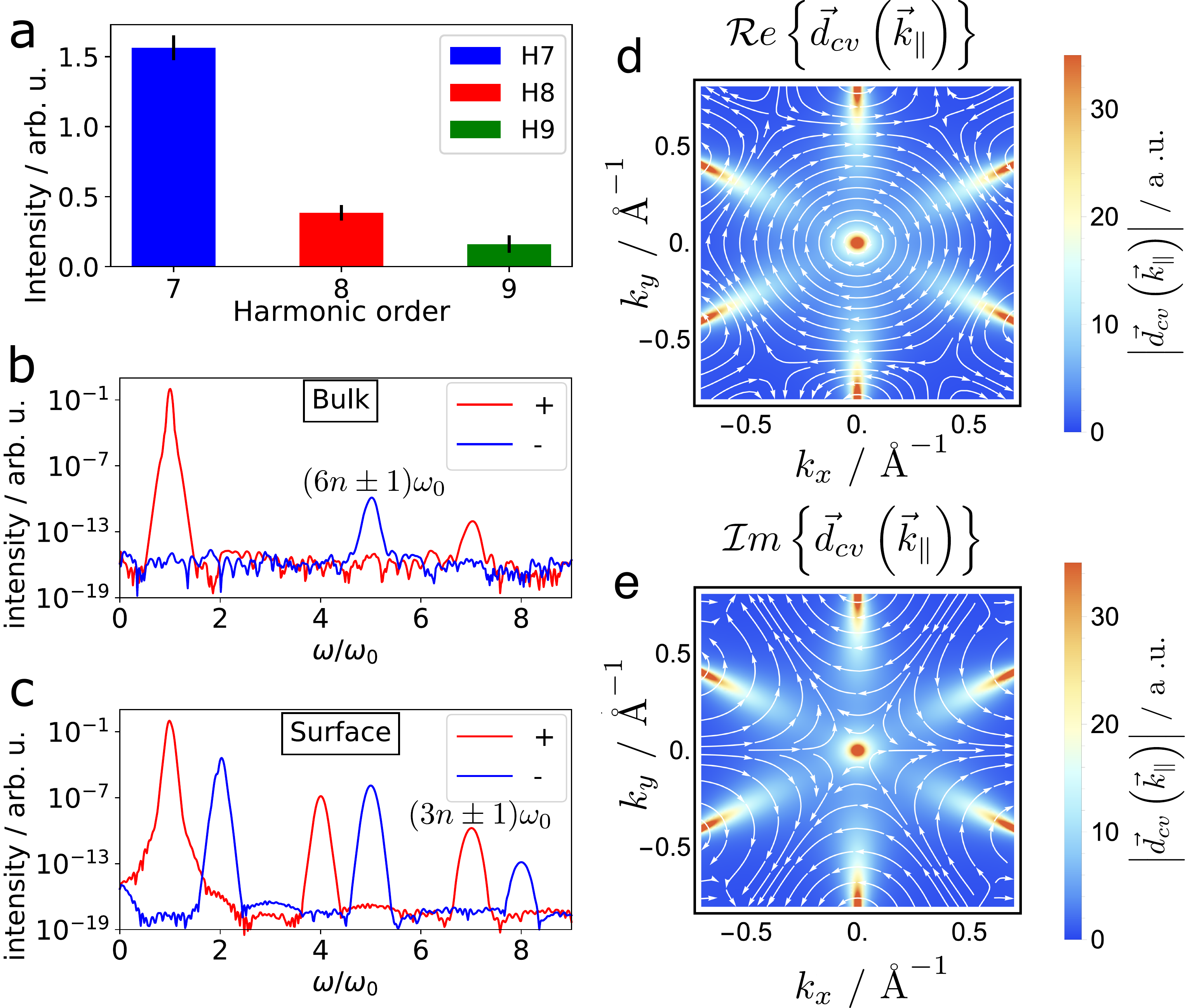}
\end{center}
\caption{\textbf{Mechanism of HHG from $\mathrm{Bi_2Se_3}$ in CPL fields}: (a) Integrated harmonic intensity generated with a CPL MIR driver, normalized with respect to the average of the signal over the entire ellipticity scan for the data presented in Fig.~\ref{fig4}~a-c.  The CPL signal is averaged over the left/right helicity contributions, and the error bars correspond to one standard deviation. Panels (b) and (c) illustrate the CPL selection rules  for  bulk and surface separately, where $(3n+1)\omega_0$ and $(6n+1)\omega_0$-orders are co-rotating (LCP, (+), red),  $(3n+2)\omega_0$ and $(6n+2)\omega_0$-orders have the opposite helicity (RCP, (-), blue) with respect to the left-circularly polarized MIR driver . (d) and (e): Dipole matrix element $\vec{d}_{cv}(\vec{k}_\parallel)$ connecting the lower and the upper part of the Dirac cone. The color quantifies the absolute magnitude; the real (panel (d)) and imaginary (panel (e)) parts of the $\hat{x}$- and the $\hat{y}$-components are visualized as a vector field (white arrows) and give rise to a helical vortex structure around the Dirac point. In panels b, c, d, and e, primary data from Figs. 3 and 5 of our related work Ref.~\cite{Baykusheva2021} has been used.}
\label{fig5}
\end{figure}

For CPL fields, selection rules for crystalline media with three-fold rotational symmetry restrict HHG from the surface to harmonics fulfilling the requirement $\omega = (3n\pm 1)\omega_0$ (with $n\in \mathcal{N}$ and $\omega_0$ being the fundamental frequency) \cite{Saito2017,Neufeld2019}. The presence of inversion symmetry in the bulk forbids even harmonic generation, hence the corresponding selection rule reads $\omega = (6n\pm 1)\omega_0$. Calculated harmonic spectra of bulk and surface states driven by a left-circularly polarized field  are displayed in Figs.~\ref{fig5}~b and~c, respectively. The experimentally observed even-harmonic emission (HO~8, cp. Fig.~\ref{fig4}~b) is only present in the surface-state spectrum (Fig.~\ref{fig4}~e). HO~9 is forbidden for both the bulk and surface states, which is also  consistent with the experimental observation ($I_\mathrm{HHG}^\mathrm{(CPL)}:I_\mathrm{HHG}^\mathrm{(LIN)} \le 0.15$), as can be discerned in Figs.~\ref{fig5}~a and \ref{fig4}~c. \\

 In our theoretical analysis, we calculate the HHG response in terms of the contributions from the bulk and the surface states as described in detail in Ref.~\cite{Baykusheva2021}. For all harmonics orders considered, the bulk contribution decreases monotonically as the ellipticity increases from $\epsilon=0$ to $|\epsilon|=1$ ( Fig.~\ref{fig4}~d, e, f, dashed lines, empty diamonds), whereas the contribution from the surface (solid curve, filled symbols) allows for significant yields for fully circular fields. For example, the theory correctly reproduces the enhancement of the harmonic yield of HO~7 at high ellipticities as well as the local maxima at $|\epsilon|\approx 0.5$ for HO~9 and the features at intermediate ellipticities for HO~8. There are, however, some discrepancies in the detailed features such as at around $\epsilon=0$ in HO~7. Although slightly unusual ellipticity dependence has been reported in other solid-state systems such as MgO~\cite{You2017} and graphene~\cite{Yoshikawa2017}, a substantial selective enhancement of the co-rotating odd-order harmonics for circularly polarized laser field represents a truly distinct feature that is reported here. Here, the origin of non-trivial ellipticity dependent response is unique to the TSSs, which is in contrast to the response from the bulk states of the same material. We now turn to elucidating links of such a markedly different ellipticity-dependent behavior of high harmonics with the physical parameters of the the surface states in Bi$_2$Se$_3$.\\

%($\lambda_\mathrm{MIR}=7.5$~$\mathrm{\mu m}$, 12-cycle Gaussian envelope, $I_0=0.0025$~TW/cm$^2$)
 Panels d and e of Fig.~\ref{fig5} display the transition dipole matrix element connecting the lower $(v)$ and the upper $(c)$ Dirac cones in reciprocal space, decomposed in terms of its real (panel d) and imaginary (panel e) components. In the region of low momenta, the highly localized and singular nature of the dipole vector forms a helical vortex-like pattern encircling the Dirac point, as illustrated by the white stream lines in Fig.~\ref{fig5}~d. At higher momenta, hexagonal ``ridge-like’’ regions of increasing dipole amplitude can be clearly distinguished. These structures are mediated by the higher-order (i.e.\ $\mathcal{O}(k_\parallel^3)$), ``warping’’ terms in the effective Hamiltonian, which cause the hexagonal deformation of the surface energy surface and also represent the counterpart of the Dresselhaus spin-orbit interaction in rhombohedral structures~\cite{Fu2009a}. These terms are responsible for the \textit{out-of-plane} spin polarization of the surface bands. By varying these specific parameters in the tight-binding model Hamiltonian, we find that the enhanced response of relatively low-order harmonics has its origin in the vortex-like pattern around the Dirac point, whereas the yield of higher-order harmonics is predominantly mediated by the parameters related to the spin-orbit coupling and out-of-plane spin polarization \cite{Baykusheva2021}.  In contrast, the transition dipole matrix elements of the bulk bands (plotted in panels~c and~d of Fig.~\ref{fig:BZ} in the SM) have negligible amplitudes in the low momentum range and exhibit annular maxima only in the momentum range where the energy separation of the inverted bands (cp. Fig.~\ref{fig:BZ}~b in the SM) is smallest. Hence, the demonstrated non-trivial response of CPL field-driven HHG can be linked to the topology of the band structure that originates in the interplay of strong spin-orbit couplings and time-reversal symmetry protection, and it underlines the potential of our technique to probe such features of electronic structure in a purely optical setting. \\

\textcolor{black}{Finally, we note that quantitative features, such as the exact positions of the maxima of HOs~8 and~9 as well as the exact enhancement ratio for HO~7 at  $|\epsilon|= 1$, are not captured correctly by our model. This is most likely due to the simplicity of our model, as we use a $4\times4$ TBM Hamiltonian and the adopted  parametrization does not reproduce the band structure of Bi$_2$Se$_3$ with quantitative accuracy, especially in the high-momentum regions of the band structure. Further, couplings between surface and bulk bands are neglected. Strictly speaking, the wavefunctions and the dispersions of the surface bands are accurate only in the low-energy limit (i.e. up to $20~\%$ of the BZ away from the zone center). Also, details of the macroscopic propagation of the emitted HHG are not taken into consideration within the current theoretical framework. Nevertheless,  our model captures the salient features of the non-linear strong-field response observed experimentally and thus can provide useful insights into the complex microscopic physics of 3D-TIs in strong laser fields.}

\section*{Conclusion and Outlook}

In conclusion, we measured an anomalous laser-ellipticity dependence of high harmonics from the prototypical topological insulator Bi$_2$Se$_3$. It is manifested in a significant enhancement of co-rotating high-harmonic orders for circularly-polarized laser fields, which is attributed to the non-trivial topology of the surface Bloch bands, in particular the dipole couplings and Berry connections close to the $\Gamma$-point and the influence of the high-order ``warping'' terms away from the zone center. Our interpretation is based on theoretical analysis combining a quantum-mechanical treatment of the HHG process in the framework of the semiconductor Bloch equations formalism with a minimal $4\times4$-tight-binding Hamiltonian for 3D topological insulators. Identification of relevant experimental observables and their relations to the topological properties of the system lays the foundation for forthcoming (and possibly time-resolved) studies of topological phase transitions based on HHG. For instance, by tracking the efficiency of high-harmonics generated by the elliptically-polarized laser fields in a  pump-probe setup one could monitor the inter-conversion between two distinct topological phases if the underlying crystalline symmetries are different, such as in the Weyl semimetal compound WTe$_2$~\cite{Sie2019}. The anomalous ellipticity behavior can be exploited to track the influence of the spin-orbit coupling in strong-field light-matter interactions by means of doping Bi$_2$Se$_3$ with lighter elements~\cite{Brahlek2012} or magnetic impurities~\cite{Chang2013} in a controlled manner. The unique strengths of the HHG-based approach is the unprecedented access to sub-cycle strong-field driven dynamics~\cite{Silva2019a}. The reported highly non-trivial response to circularly polarized driving fields  provides novel opportunities for the study of chiral phenomena such as helical currents on the surfaces of  topological materials, with implications for both attosecond metrology and light-wave-driven nanoelectronics~\cite{Garg2016,Reimann2018,OliaeiMotlagh2018, Hafez2018}. A forthcoming theoretical challenge consists in the development of a methodology for the extraction of topological invariants \cite{Chacon2018,Bauer2018,Silva2019a} from experimentally accessible observables.

\section*{Materials and Methods}
\subsection*{Experimental setup}
Femtosecond pulses in the MIR spectral range (5--\textcolor{black}{10}~$\mathrm{\mu m}$) are generated in a setup (see Fig.~\ref{fig:exp} in the Supplementary Material) based on a 1~kHz regenerative amplifier and optical parametric amplification followed by di\-fference-frequency generation in GaSe. The output of the Ti:Sa amplifier (Coherent, 7~mJ, 45~fs, with a center wavelength of $\sim$790~nm) is used to pump a commercial optical parametric amplifier (TOPAS-HE, Light Conversion) in order to generate signal and idler pulses of \mbox{$\sim$48--55 fs} typical duration with central wavelengths tuned between 1415--1435~nm and 1715--1820~nm, respectively. Non-collinear type-I-phase-matched  difference-frequency-mixing of the OPA output in a 500~$\mathrm{\mu m}$ thick, $z$-cut GaSe crystal (Eksma optics) yields MIR pulses tunable in the range 5--10~$\mathrm{\mu m}$ reaching typical energies of 80~$\mathrm{\mu J}$ at $7.5$~$\mathrm{\mu m}$. A 3~mm Ge filter is used to separate the MIR from the residual signal and idler beams. A thin Ge (0.5~mm) Brewster plate is used to clean the polarization of the horizontally polarized DFG output. The frequency of the driving MIR field is quantified using Michelson interferometry, whereas the pulse duration is estimated using SHG-autocorrelation measurements in a AgGaS$_2$ crystal. The findings presented in the main text have been obtained with  \textcolor{black}{MIR pulses tunable in the range from $5$~$\mathrm{\mu m}$ to $10$~$\mathrm{\mu m}$} and $\approx 300$~fs in duration. The parameters for the various datasets presented in the main text and in the SM are summarised in Tab.~S1 in the SM.\\

The MIR is focused onto the sample using a silver-coated ($90^\circ$) off-axis parabolic mirror of 10~cm  focal length.  In practice, the sample is positioned few mm away from the focal point, the resulting combination of a loose-focusing geometry and a large beam spot on the sample surface minimizes damage while ensuring sampling over a larger surface area. The beam waist ($1/e^2$) at the sample position is 300~$\mathrm{\mu m}$, as determined with a knife-edge scan measurement. The polarization state of the driving field is varied with the aid of tunable half- (HWP, $\lambda/2$) and quarter-wave (QWP, $\lambda/4$) plates designed for the spectral region 1--19~$\mathrm{\mu m}$ (5~mm CdSe, Alphalas). The ellipticity scan measurements presented in Fig.~4 of the main text as well as Fig.~\ref{fig:ellscan_60} in the SM were performed by rotating the quarter-wave plate while simultaneously rotating the crystal by the same angle in order to keep the major axis of the polarization ellipse aligned to the selected high-symmetry direction. The design of the tunable waveplate allows for pre-compensating the phase delay introduced by the different Fresnel coefficients of the S/P components as well as the distortions introduced by the parabolic mirror by adjusting the tilt of the waveplate. The polarization of the MIR in the focal plane was confirmed to be nearly circular, with ellipticities reaching 0.95 (cp. Fig.~\ref{fig:ell_supp}~a in the SM). \\

The harmonic emission is re-collimated, spectrally dispersed in a monochromator (Acton series, Princeton instruments) and imaged onto a CCD camera (Andor DO440-BN). All experimental measurements have been conducted at ambient conditions.\\

Due to limitations imposed by the instrument response, lower-order (than HO~7) harmonic orders lie outside of the detection range.
\subsection*{Sample growth and characterization}\label{sec:sample}
Thin Bi$_2$Se$_3$ films of \textcolor{black}{sub-100-nm thickness} are grown by molecular beam epitaxy (MBE) on lattice-matched BaF$_2$ (111) substrates. BaF$_2$ is a wide-band gap insulator (band gap of $\sim11$~eV~\cite{Rubloff1972}), transparent and optically isotropic, making it particularly suitable for optical spectroscopy. BaF$_2$ substrates (1~cm${}^2$ $\times$ 0.5~mm, polished on both sides) were cleaned in boiling acetone and ethanol and subsequently baked at $\sim800^\circ$~C for 30~minutes. The substrates were held at $\sim225^\circ$~C during growth and annealing. A Se:Bi flux ratio of ~7 was used during the growth, as calibrated by the in-situ crystal monitor. The films were annealed for 90~minutes in Se flux in order to reduce Se vacancy formation. The films  were capped by a 2~nm protective Te layer. The orientation of the Bi$_2$Se$_3$ structure (along the $c$-axis) as well as the high crystallinity of the films were confirmed by standard laboratory X-ray diffraction measurements. \\

\section*{Acknowledgments}
\noindent
We acknowledge Marc Welch for technical assistance. Theoretical calculations have been performed using the Sherlock HPC cluster at Stanford University as well as using computational resources provided by the Fundaci\'{o}n P\'{u}blica Galega Centro Tecnol\'{o}xico de Supercomputaci\'{o}n de Galicia (CESGA).

\section*{Funding}
\noindent
This work was primarily supported by the US Department of Energy, Office of Science, Basic Energy Sciences, Chemical Sciences, Geosciences, and Biosciences Division through the Early Career Research Program.  D.B. gratefully acknowledges support from the Swiss National Science Foundation (SNSF) through project No: P2EZP2\_184255. A.C. acknowledges financial support by the Max Planck POSTECH/KOREA Research Initiative Program (Grant No.~2016K1A4A4A0192202) through the National Research Foundation of Korea (NRF) funded by the Ministry of Science, ICT and Future Planning, Korea Institute for Advancement of Technology (KIAT) grant funded by the Korea  Government (MOTIE) (P0008763, The Competency Development Program for Industry Specialists) and LANL LDRD project. 
\section*{Author contributions}
\noindent
D.B., S.G., and A.C. conceived the project. D.B. and J.L. constructed the experimental setup. D.B. recorded and analyzed the data. D.B. and A.C. developed the theoretical model and performed the numerical calculations. C.R.R. and T.P.B. synthesized and characterized the samples.  H.S., J.A.S., P.S.K., T.F.H., and D.A.R. contributed in interpreting data. All authors discussed the manuscript. 
\section*{Competing interests}
\noindent
The authors declare no competing interests.
\section*{Data and materials availability}
\noindent
All data needed to evaluate the conclusions in the paper are present in the paper and/or the Supplementary Materials. Raw data sets used for generating Figures 1-5 as well as the additional experimental results shown in the supplementary material will be made available on Zenodo.

\section*{Supplementary materials}
\noindent
Supplementary Text\\
Figs. S1 to S6\\
Table S1 \\

\bibliography{Bi2Se3_HHG}

\end{document}